\documentclass[sigconf,screen]{acmart}

\AtBeginDocument{%
  \providecommand\BibTeX{{%
    \normalfont B\kern-0.5em{\scshape i\kern-0.25em b}\kern-0.8em\TeX}}}

\usepackage{algorithm}

\usepackage[noend]{algpseudocode}

\usepackage{xcolor}
\usepackage{pdfcomment}
\usepackage{microtype} 
\newcommand{\methodname}{MetaIC}

\AtBeginDocument{%
  \providecommand\BibTeX{{%
    \normalfont B\kern-0.5em{\scshape i\kern-0.25em b}\kern-0.8em\TeX}}}

\setcopyright{acmcopyright}
\acmPrice{15.00}
\acmDOI{10.1145/3533767.3534389}
\acmYear{2022}
\copyrightyear{2022}
\acmSubmissionID{issta22main-p196-p}
\acmISBN{978-1-4503-9379-9/22/07}
\acmConference[ISSTA '22]{Proceedings of the 31st ACM SIGSOFT International Symposium on Software Testing and Analysis}{July 18--22, 2022}{Virtual, South Korea}
\acmBooktitle{Proceedings of the 31st ACM SIGSOFT International Symposium on Software Testing and Analysis (ISSTA '22), July 18--22, 2022, Virtual, South Korea}

\begin{document}

\title{Automated Testing of Image Captioning Systems}
\author{Boxi Yu}
\email{221049024@link.cuhk.edu.cn}
\affiliation{
  \institution{The Chinese University of Hong Kong, Shenzhen}
  \city{Shenzhen}
  \state{Guangdong}
  \country{China}
}

\author{Zhiqing Zhong}
\email{cheehingchung@gmail.com}
\affiliation{%
  \institution{South China University of Technology}
  \city{Guangzhou}
  \state{Guangdong}
  \country{China}
}

\author{Xinran Qin}
\email{csqinxinran@mail.scut.edu.cn}
\affiliation{%
  \institution{South China University of Technology}
  \city{Guangzhou}
  \state{Guangdong}
  \country{China}
}

\author{Jiayi Yao}
\email{120040070@link.cuhk.edu.cn}
\affiliation{%
  \institution{The Chinese University of Hong Kong, Shenzhen}
  \city{Shenzhen}
  \state{Guangdong}
  \country{China}
}

\author{Yuancheng Wang}
\email{119010319@link.cuhk.edu.cn}
\affiliation{%
  \institution{The Chinese University of Hong Kong, Shenzhen}
  \city{Shenzhen}
  \state{Guangdong}
  \country{China}
}

\author{Pinjia He}
\authornote{Corresponding author.}
\email{hepinjia@cuhk.edu.cn}
\affiliation{%
  \institution{The Chinese University of Hong Kong, Shenzhen}
  \city{Shenzhen}
  \state{Guangdong}
  \country{China}
}



\begin{abstract}
Image captioning (IC) systems, which automatically generate a text description of the salient objects in an image (real or synthetic), have seen great progress over the past few years due to the development of deep neural networks. IC plays an indispensable role in human society, for example, labeling massive photos for scientific studies and assisting visually-impaired people in perceiving the world. However, even the top-notch IC systems, such as Microsoft Azure Cognitive Services and IBM Image Caption Generator, may return incorrect results, leading to the omission of important objects, deep misunderstanding, and threats to personal safety. 

To address this problem, we propose {\methodname}, the \textit{first} metamorphic testing approach to validate IC systems.
Our core idea is that the object names should exhibit directional changes after object insertion.
Specifically, {\methodname} (1) extracts objects from existing images to construct an object corpus; (2) inserts an object into an image via novel object resizing and location tuning algorithms; and (3) reports image pairs whose captions do not exhibit differences in an expected way.
In our evaluation, we use {\methodname} to test one widely-adopted image captioning API and five state-of-the-art (SOTA) image captioning models. Using 1,000 seeds, {\methodname} successfully reports 16,825 erroneous issues with high precision (84.9\%-98.4\%). There are three kinds of errors: misclassification, omission, and incorrect quantity.  
We visualize the errors reported by {\methodname}, which shows that flexible overlapping setting facilitates IC testing by increasing and diversifying the reported errors.
In addition, {\methodname} can be further generalized to detect label errors in the training dataset, which has successfully detected 151 incorrect labels in MS COCO Caption, a standard dataset in image captioning.
\end{abstract}

\ccsdesc[500]{Software and its engineering~Software testing and debugging}




\keywords{Metamorphic testing, testing, image captioning, AI software}


\maketitle


\section{Introduction}\label{sec:intro}


\begin{figure*}[ht]
		\centering
		\includegraphics[width=1.0\linewidth]{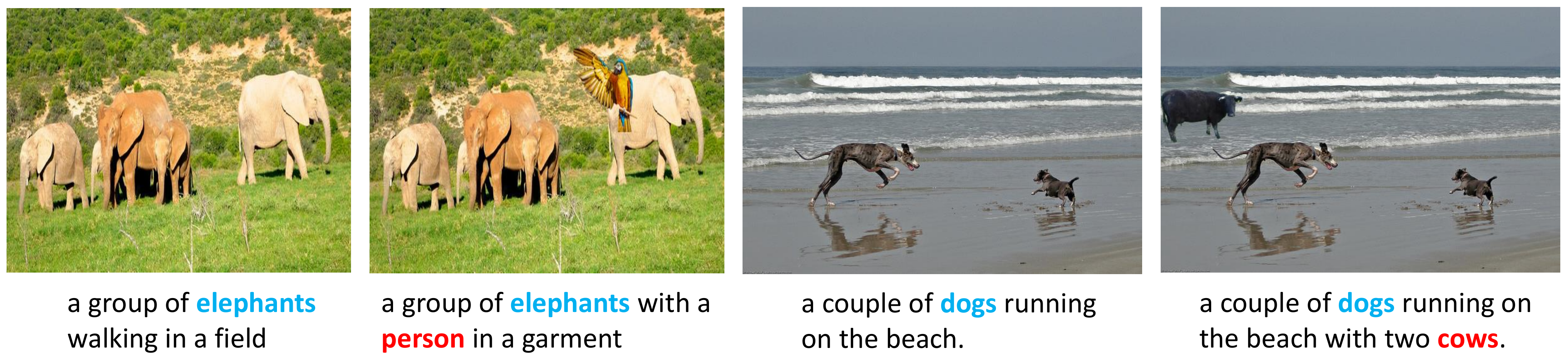}
		\caption{ Examples of background images, generated images, and the corresponding captions. Words of salient objects are marked in blue. Words indicating the violation of {\methodname}'s MRs are marked in red.}
		\label{fig:how_MRs_work}
\end{figure*}

Image captioning (IC) systems aim to automatically generate a brief depiction of the salient objects in an image (real or synthetic). 
In recent years, the performance of IC systems~\cite{karpathy2015deep, vinyals_show_2015, xu_show_2015, li_oscar_2020, hu_vivo_2020, zhang_vinvl_2021} have improved significantly due to the rapid development of the underlying deep neural networks, such as convolutional neural networks (CNN)~\cite{ren2015faster, he2016deep} for image feature extraction and language models~\cite{hochreiter1997long,devlin2018bert} for caption generation based on these features.
The SOTA IC models have reached or even surpassed human-level performance (\textit{e.g.}, VIVO~\cite{hu_vivo_2020}) in terms of CIDEr score~\cite{vedantam2015cider}.
Thus, IC systems have been increasingly integrated into our daily lives.
Typical examples include generating captions for massive social media photos~\cite{ic2socialnetwork}, visually-impaired people's artificial intelligence (AI) assistants~\cite{ic2impariment, ic2impariment_2}, and the acceleration of various manual image labeling tasks~\cite{ic2labeldata}.
Since 2016, the major IT companies (\textit{e.g.}, Google~\cite{ic2labeldata}, Microsoft~\cite{miocrosoft_azure_api}, and IBM~\cite{ibm_imgcap_api}) have released and continuously improved their own IC systems.
The leading company of Geographic Information System (GIS), Ersi, has integrated image captioning into its famous GIS framework (\textit{e.g.}, Arcgis\footnote{https://developers.arcgis.com/python/guide/how-image-captioning-works/}) to generate the captions for remote sensing images. 


Despite its wide adoption in various applications, modern IC system could return incorrect captions due to the complexity of deep neural networks and the labeling errors in training datasets, leading to the omission of important objects, deep misunderstanding, and even threats to personal safety~\cite{news_1_ai_nodate,ic2impariment,schroeder_microsoft_2016}. According to the "report on vision"~\cite{organization_world_2019}, more than 2.2 billion people around the world have vision impairment and they are expected to benefit greatly from AI-assistants in which IC is often a major component~\cite{ic2impariment,ic2impariment_2}. However, the embedded IC systems sometimes fail to generate the correct captions~\cite{news_1_ai_nodate}, which make the AI-assistants return misleading message to the visually-impaired users, posing threat to their personal safety. An image of the former U.S. president Barach Obama and his wife was captioned "a man in suit and tie talking on a cell phone"~\cite{schroeder_microsoft_2016}, leading to misunderstanding and potential negative social impact. Thus, assuring the reliability of IC systems is an important endeavor.

There remains a dearth of automated testing methods for IC systems because the problem is quite challenging.
First, modern IC systems are approaching human-level performance on the existing human-labeled test sets. Thus, test sets of high quality are lacking. 
Second, traditional code-based testing approaches~\cite{romdhana2021cosmo, viticchie2016assessment, ceccato2014family} are not suitable for testing IC systems because the logics of IC systems are mainly encoded in the underlying neural networks with millions of complex parameters, rather the source code. Third, existing testing approaches for computer vision (CV) tasks~\cite{wang2020metamorphic, tian2020testing} typically assume simpler output formats (\textit{e.g.}, class labels), which cannot work well in testing IC systems whose output (\textit{i.e.}, a sentence) is significantly more complex; while existing testing approaches for natural language processing (NLP)~\cite{sun2020automatic,he2020structure,he2021testing} tasks heavily rely on the perturbation of chars or words in the input, which is infeasible for IC systems whose input are images.  

To tackle these challenges, we introduce {\methodname}, a simple, widely-applicable metamorphic testing methodology for validating IC systems. The input of {\methodname} is a list of unlabeled images, while its output is a list of suspicious issues, where each suspicious issue contains a pair of images and their captions.   
The core idea of {\methodname} is that the object names should exhibit directional changes after object insertion. We realize this idea by two metamorphic relations (MRs): (1) \textit{object appearance}: only the original objects and the inserted objects should appear in the caption of the generated image; (2) \textit{singular-plural form}: both the original objects and the inserted objects should be illustrated by words of a proper singular-plural form. 
Fig.~\ref{fig:how_MRs_work} presents two suspicious issues returned by our approach.
The first issue was reported because the inserted bird was captioned into a "person", violating the first MR; while the second issue was reported because the noun "cow" was not in singular form, violating the second MR. Specifically, {\methodname} automatically extracts objects from images by Yolact++~\cite{bolya2020yolact++}, an advanced real-time image segmentation technique. The extracted object will be inserted into randomly selected images (\textit{i.e.}, background images). This is a challenging step because the inserted object should avoid affecting the saliency of the existing objects, which has been tackled by our novel object resizing and location tuning algorithms.
The background image and the newly generated image form an image pair. If the captions of this image pair returned by the IC system violate any of the MRs, the image pair and the corresponding captions will be reported as a suspicious issue. 

We apply {\methodname} to test one paid IC service, \textit{i.e}., Microsoft Azure Cognitive Services~\cite{miocrosoft_azure_api} and five popular IC models, \textit{i.e.}, Show, Attend and Tell~\cite{xu_show_2015}, $Oscar_B$, $Oscar_L$~\cite{li_oscar_2020}, $VinVL_B$, $VinVL_L$~\cite{zhang_vinvl_2021}. 
{\methodname} successfully reports 16,825 erroneous issues in total with high precision of 84.9\%-98.4\%, revealing 17,380 captioning errors. The errors include misclassification, omission, and incorrect quantity. To better understand the reported errors, we visualize the attention masks of the object, which helps us explore the potential root causes behind. Furthermore, we adapt {\methodname} to detect labeling errors in MS COCO Caption, a widely-used standard IC dataset, which reports 151 incorrect labels in the training set, demonstrating {\methodname}'s wide applicability. The source code is also released for reuse~\cite{metaic}. In summary, this paper makes the following main contributions:
\begin{itemize}
    \item It introduces {\methodname}, the \textit{first} metamorphic testing methodology for validating image captioning systems.
    \item  It describes a practical implementation of {\methodname} by adapting Yolact++~\cite{bolya2020yolact++} to extract objects and developing a new object insertion technique that allows flexible insertion position and overlapping area.
    \item It presents the evaluation of {\methodname} that successfully reports 16,825 erroneous issues in one industrial IC service and five SOTA IC models with high precision, and find a total number of 17,380 captioning errors.
    \item It discusses the error categories found by {\methodname}, the exploration of root causes, and the potential of adapting {\methodname} to detect labeling errors in widely-used image captioning datasets.
\end{itemize}

\begin{figure}[t]
		\centering
		\includegraphics[width=0.95\linewidth]{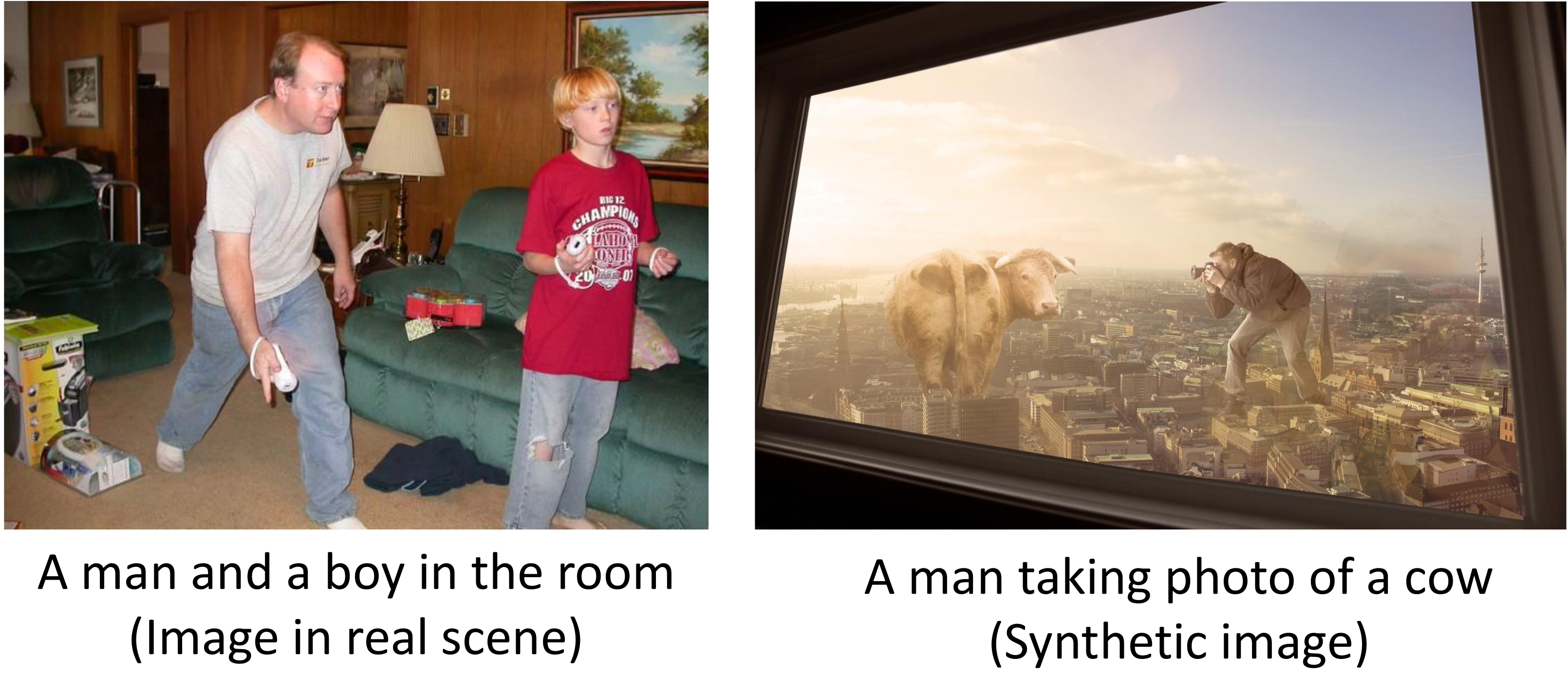}
		\caption{ A real image, a synthetic image, and their captions. }
		\label{fig:real_syn}
\end{figure}

\section{Preliminaries}\label{sec:pre}
In this section, we first explain the basic concepts that have been used multiple times in the paper. Then we introduce two mainstream image captioning models: CNN-RNN-Based IC and Vision-Language Pre-Training IC.


\subsection{Basic Concepts}

Image captioning outputs a depiction of the \textit{salient objects} in a given image (real or synthetic), where salient objects refer to the most noticeable object(s) in the image; and \textit{saliency} reflects whether an object is salient or not. 
For example, in Fig.~\ref{fig:real_syn}, the two people are salient objects in the real-scene image, and the man and the cow are salient objects in the synthetic image. Each object has an \textit{object class}, such as "elephant" and "dog" in Fig.~\ref{fig:how_MRs_work}. In a caption returned by IC software, a salient object is typically illustrated by a noun, whose \textit{singular-plural form} indicates whether the word is a singular noun or a plural noun.

\begin{figure*}[ht]
		\centering
		\includegraphics[width=1.0\linewidth]{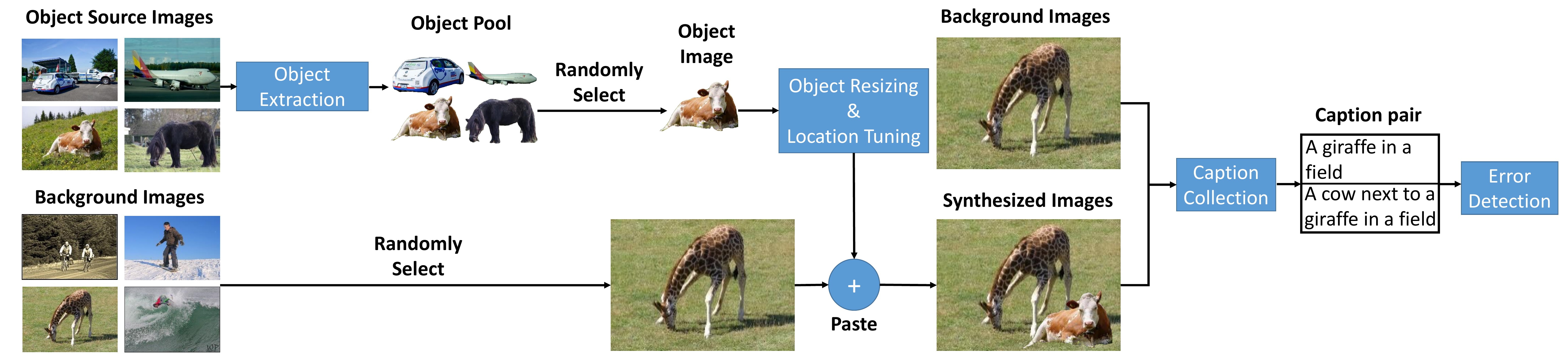}
		\caption{Overview of {\methodname}.}
		\label{fig:framework}
\end{figure*}

In this paper, we test one IC service and five IC models: Microsoft Azure API~\cite{miocrosoft_azure_api}, Attention~\cite{xu_show_2015}, $Oscar_B$, $Oscar_L$~\cite{devlin2018bert}, $VinVL_B$, $VinVL_L$~\cite{zhang_vinvl_2021}. Specifically, 
Microsoft Azure is short for Microsoft Azure Cognitive services~\cite{miocrosoft_azure_api} and Attention is short for Show, Attend and Tell~\cite{xu_show_2015}. $Oscar_B$ and $VinVL_B$ refer to the IC models adopting the base version of BERT~\cite{devlin2018bert}, while $Oscar_L$ and $VinVL_L$ adopt the large version.

\subsection{Modern Image Captioning Systems}


Recently, there are two main lines of IC research: CNN-RNN-Based IC~\cite{xu_show_2015, vinyals_show_2015} and Vision-Language Pre-Training (VLP) IC~\cite{zhou2020unified, hu_vivo_2020, li_oscar_2020, zhang_vinvl_2021}. They both adopt a two-stage framework, using CNNs for feature extraction in the first stage and sequence models caption generation in the second stage. Differently, VLP IC adopts pre-trained models in the second stage, which makes them more efficient and accurate than CNN-RNN-Based IC. In particular, the SOTA VLP IC model VIVO~\cite{hu_vivo_2020} claims that it surpasses human in CIDER score~\cite{vedantam2015cider}. In the following, we elaborate more on these two lines of research.

\subsubsection{CNN-RNN-Based ICs}

CNN-RNN-Based IC uses a CNN as the feature extractor and an RNN to generate the captions.  
Specifically, CNN can produce rich representation of an image by embedding it into a fixed-length vector and the representation has been used in various CV tasks~\cite{sermanet2013overfeat}.
Inspired by the great successes of sequence generation in machine translation~\cite{bahdanau2014neural, cho2014learning, sutskever2014sequence}, RNN models~\cite{karpathy2015deep,xu_show_2015,vinyals_show_2015} have been adopted in caption generation in CNN-RNN-Based IC. Attention~\cite{xu_show_2015} first introduces an attention mechanism to image captioning and visualize how the model attends on the salient objects in an image during captioning. 
By using a lower convolutional layer rather than fully connected layers in the encoder, the RNN decoder of this method can focus on the most important parts of the image by calculating weights from the feature vectors.
As for the second stage, it adopts LSTM~\cite{hochreiter1997long} for caption generation.


\subsubsection{Vision-Language Pre-Training ICs}

VLP ICs also adopts a two-stage pipeline. Differently, they employ more advanced CNNs in the first stage and BERT~\cite{devlin2018bert} in the second stage. With a multi-layer bidirectional Transformer~\cite{vaswani2017attention} as its architecture, BERT uses masked language models to enable pre-trained deep bidirectional representations, which effectively alleviate the extensive manual effort on building task-specific architectures. 

Oscar~\cite{li_oscar_2020} first adopts object tags detected in images as anchor points, which strengthens the learning of semantic alignments between images and texts and significantly enhance its ability to learn the cross-modal representations. 
Recently, Zhang \textit{et al.}~\cite{zhang_vinvl_2021} propose VinVL, which uses a CNN specially tailored for vision-language tasks. This CNN is pre-trained on a much larger text-image corpora, thus it can capture much more abundant visual features accurately, outperforming other models.

\section{Approach and Implementation}\label{sec:method}


This section introduces {\methodname} and its implementation details. Inspired by the goal of IC that the salient objects in an image should be captioned properly, the core idea of {\methodname} is: the object names should exhibit directional changes after object insertion.
The input of {\methodname} is a list of unlabeled images and the output is a list of suspicious issues, where each issue contains the original image, a generated image with an inserted object, and their corresponding captions returned by the IC software under test. Fig.~\ref{fig:framework} illustrates the overview of {\methodname}, which carries out the following four steps:
\begin{enumerate}
    \item \textit{Object extraction}. We extract salient objects from an existing dataset by an instance segmentation algorithm, which collectively form an object pool.
    \item \textit{Object insertion}. For each unlabeled image (\textit{i.e.}, background image), we randomly select an image (\textit{i.e.}, object image) from the object pool and insert it into the background image via novel object resizing and location tuning algorithms.  
    \item \textit{Caption collection}. We collect the captions for the background image and the generated image from the IC system under test.
    \item \textit{Error detection}. We focus on the key constituents in the two captions and report a suspicious issue if the captions violate at least one of our MRs.
\end{enumerate}

\subsection{Object Extraction}

To realize object insertion, we need to prepare a pool of object images. Thus, the first step of {\methodname} is to extract object images from an image dataset. 
Although object extraction has long been a challenging task, deep learning-based object segmentation models~\cite{bolya2019yolact, bolya2020yolact++, he2017mask, li2017fully} provide a feasible solution because these models can extract high-quality objects efficiently. 

\begin{figure}[ht]
		\centering
		\includegraphics[width=1.0\linewidth]{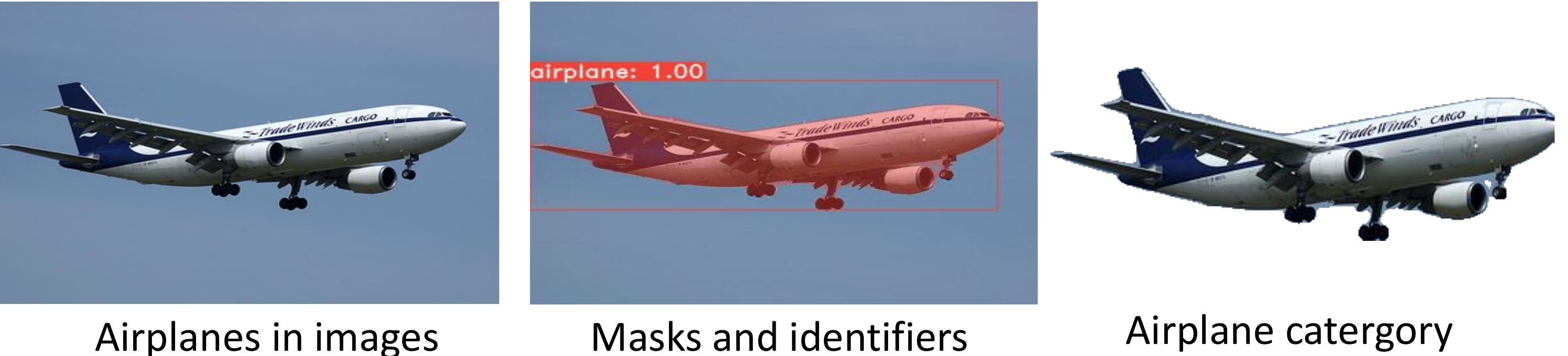}
		\caption{An example of object extraction via Yolact++.}
		\label{fig:how_yolact++_works}
\end{figure}

In our implementation, we adopt Yolact++~\cite{bolya2020yolact++}, the SOTA real-time object segmentation method because it achieves decent performance in terms of both precision~\cite{he2017mask, li2017fully} and efficiency~\cite{uhrig2018box2pix,bolya2019yolact,bolya2020yolact++}. Yolact++ predicts mask prototypes and per-instance mask coefficients in parallel, and linearly combines them to form the final instance masks. 
Thus, Yolact++ model can achieve 34.1 mAP on MSCOCO~\cite{lin2014microsoft} at 33.5 fps while keeping the close performance of the SOTA approaches. As shown in Fig.~\ref{fig:how_yolact++_works}, Yolact++ generates the masks and identifiers of the airplanes in the images, based on which we can extract the objects accordingly and put them into a certain category (\textit{e.g.}, airplane).

\subsection{Object Insertion}
Once we have constructed an object pool, we randomly select an object image (\textit{e.g.}, the cow in Fig.~\ref{fig:framework}) from the pool and insert it into a random background image. In this process, we need to address three main issues: (1) how to resize the object image to avoid being too big or too small compared with the salient objects in the background image; (2) how to find a suitable location for insertion; and (3) how to allow reasonable overlap between the object image and the original objects in the background image. In the following, we will introduce our specially-design object resizing and location tuning algorithm and how these algorithms provide us with the flexibility of overlapping area configuration.

\subsubsection{Object resizing.}
\begin{figure}[ht]
		\centering
		\includegraphics[width=0.9\linewidth]{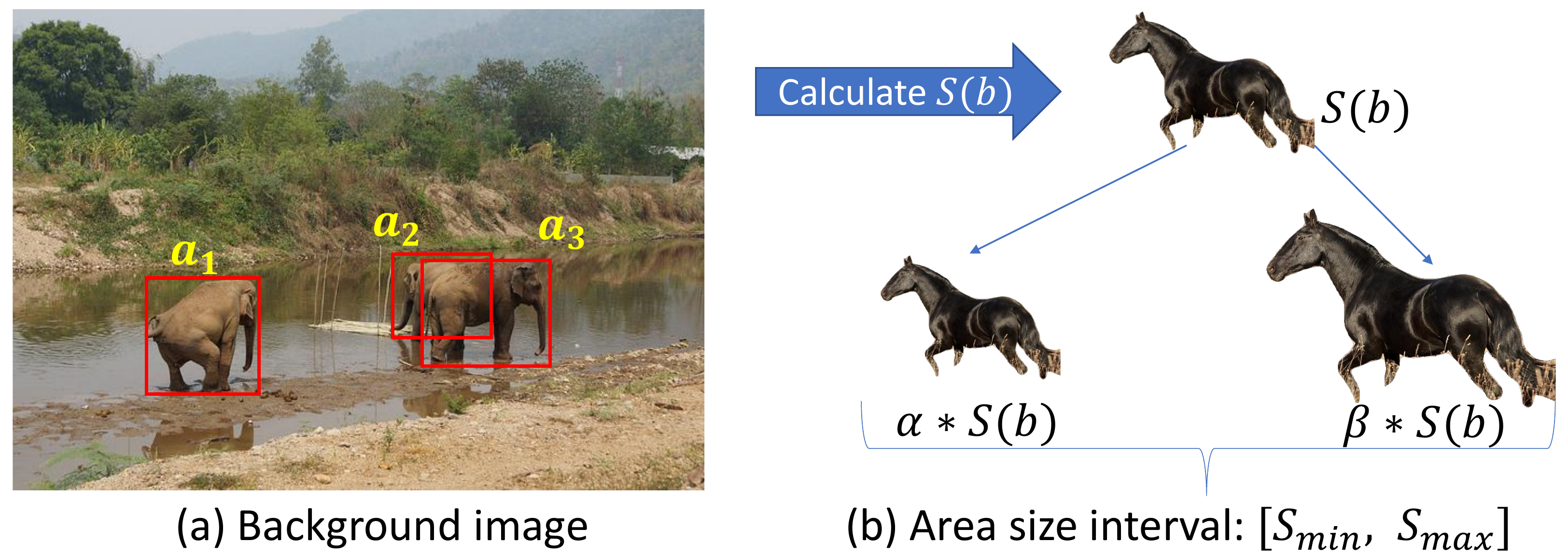}
		\caption{An example of object resizing.}
		\label{fig:resizing_show}
\end{figure}

The goal of object resizing is to resize the object image from the original size $(h, w)$ to a new size $(h', w')$, where $h$ and $h'$ refer to their heights and $w$ and $w'$ refer to their widths. {\methodname} is based on the assumption that the inserted object is a salient object and it should not affect the saliency of the objects in the background image. Thus, the resized object image should not be too small nor too big. 
Our object resizing algorithm is illustrated by Fig.~\ref{fig:resizing_show}.
To figure out the feasible $(h', w')$, we first determine the area size $s\approx h'\times w'$ by randomly selecting a value from range $[S_{min}, S_{max}]$, where $S_{min}$ and $S_{max}$ are calculated as follows:
\begin{equation}
    \begin{aligned}
    S_{min}& = \alpha*S(b) \\
    S_{max}& = \beta*S(b) \\
    \end{aligned}
\end{equation}
$\alpha$ and $\beta$ are hyper-parameters that we manually set based on empirical experience; $S(b)$ is a normalized value that reflects the area size of the objects in the background image. Specifically, $S(b)$ is calculated by $Softmax$ function as follows:

\begin{equation}
    \begin{aligned}
    S(b)& = \sum_{i=1}^{N}{ ( Softmax( \frac{a_i}{s_b} )*a_i} ) \\
    Softmax(x_i)& = \frac{e^{x_i}}{\sum_{j=1}^{N}e^{x_j}} \\
    \end{aligned}
\end{equation}
where $a_i$ is the area size of the $i$-th original objects in the background image $b$; $s_b$ is the area size of $b$; $N$ is the number of objects in $b$.  
After randomly selecting an area size $s$ from the interval $[S_{min}, S_{max}]$, we calculate
$(h', w')$ as follows:
\begin{equation}
    \begin{aligned}
    (h', w')& = (h*\sqrt{\frac{s}{h*w}}, w*\sqrt{\frac{s}{h*w}} ) \\
    \end{aligned}
\end{equation}

We denote $a_{max}$ as the area size of the largest object in $b$. Empirically when $\frac{a_{max}}{s_b}$ < $40\%$, we set $\alpha$, $\beta$ as 0.8 and 1.3, respectively; otherwise, we set $\alpha$, $\beta$ as 0.1 and 0.37, respectively.

\subsubsection{Location tuning.}
After resizing the object image, we need to select suitable locations in the background image for object insertion. {\methodname} provides a flexible setting on the overlapping ratio between the inserted object and objects in the background image. Specifically, {\methodname} can generate $n$ images with different overlapping ratios, which are randomly selected from $n$ ratio intervals given a maximum overlapping ratio $ratio_{max}$. The intervals for the overlapping ratios are calculated as follow:
\begin{equation}
    interval_i = 
    \begin{cases}
    [0] & \text { if } i = 0 \\ 
    ( \frac{ratio_{max}}{n-1} * (i-1), \frac{ratio_{max}}{n-1} * (i)  ] & \text { if } i>0
    \end{cases}
\end{equation}

For example, if we set $ratio_{max}=0.45$ and $n=4$, we will have four intervals: \textit{i.e.}, [0], (0, 0.15], (0.15, 0.3], (0.3, 0.45]), where "[0]" indicates that the inserted object should not overlap with any of the objects in the background image.
The overlapping ratio between two objects is the overlapping ratio between the bounding boxes of the inserted object and that of the objects in the background image. Specifically, the overlapping ratio $O_j$ is defined as:
\begin{equation}
    O_j = \frac{A_j\cap A_{obj}}{A_j},
\end{equation}
where $A_j$ is the region of the $j$-th object in the background image, $A_{obj}$ is the region of the inserted object $obj$. 

For each of the ratio intervals, {\methodname} generates one synthetic image by selecting a location $(x, y)$ for object image insertion, leading to overlapping ratio $O_j$ within the interval for all the objects. 
To ensure the overlapping ratio while keeping the saliency of the original objects, {\methodname} requires: (1) $O_j$ should be in the ratio interval if the $j$-th object is the largest object in the background image (Rule $R_1$); (2) otherwise, $O_j$ should be less than or equal to the upper bound of the ratio interval Rule $R_2$.
It is challenging to select suitable positions that obeys rules $R_1$ and $R_2$ in an efficient manner because there could be multiple objects in a background image, where each poses an overlapping constraint according to the rules.
To tackle this challenge, we develop a two-step algorithm. We first search a suitable coordinate for object insertion that allows an overlapping ratio in $interval_{n-1}$, \textit{e.g.}, (0.3, 0.45] when $ratio_{max}=0.45$ and $n=4$. Then we searches the suitable positions for other ratio intervals only along a line instead of all possible positions. Alg.~\ref{alg:location} presents pseudo-code.

\textbf{Step 1}: We randomly choose a coordinate $(x_{n-1}, y_{n-1})$ in the background image for object insertion for $interval_{n-1}$ , and check if it obeys $R_1$ and $R_2$ (line 3-7).
If not, {\methodname} calculates a new image size using our image resizing technique, generates a new coordinate $(x, y)$ randomly, and check its compliance with $R_1$ and $R_2$ again. We keep generating new image sizes and coordinates until they obey $R_1$ and $R_2$ or we try $c_1$ times, where $c_1$ is a pre-defined threshold (line 7-11). 

\begin{figure}[t]
		\centering
		\includegraphics[width=1.0\linewidth]{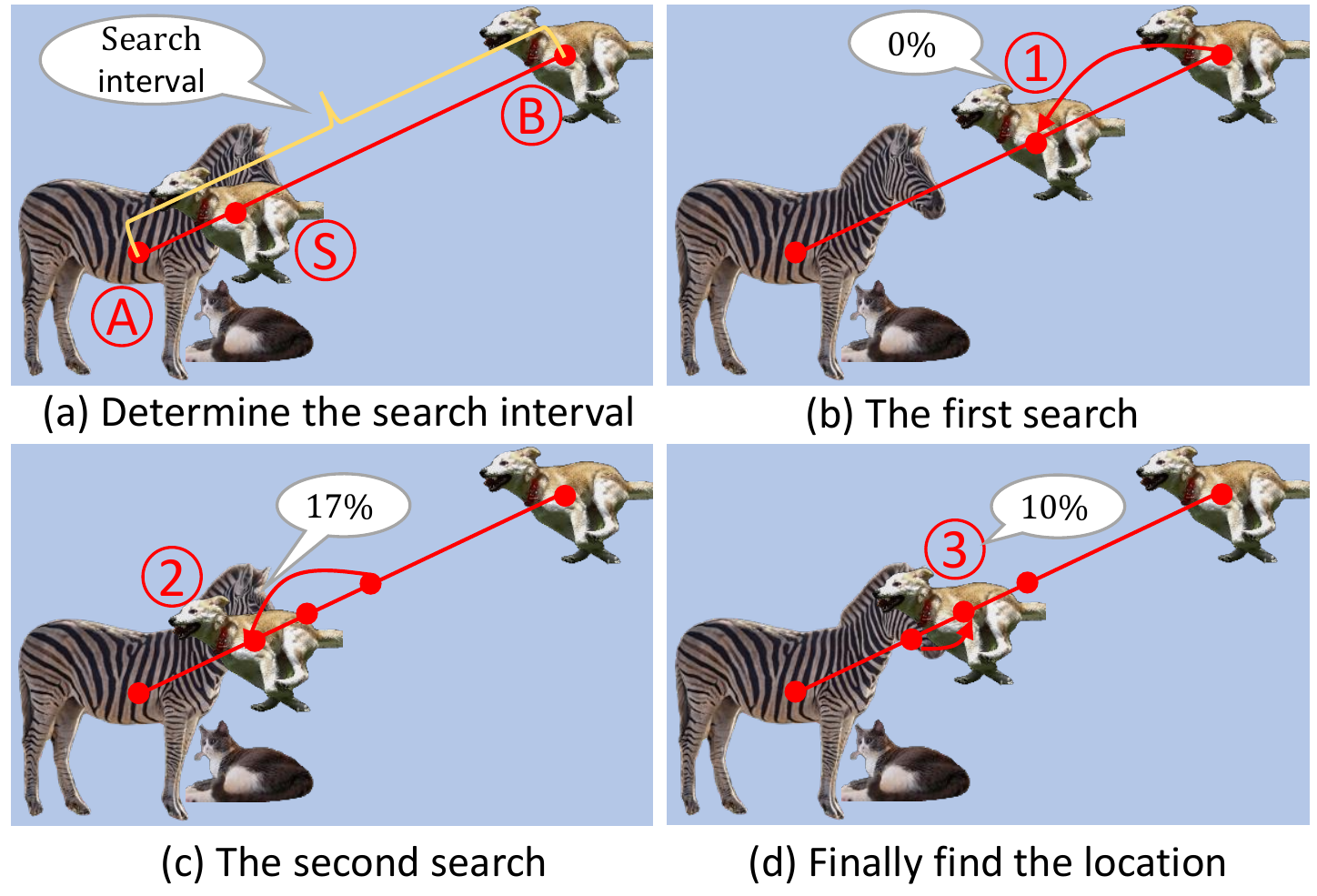}
		\caption{An example of our location technique.}
		\label{fig:insert_example}
\end{figure}

\textbf{Step 2}: After calculating the coordinate (\textit{e.g.}, point $S$ in Fig.~\ref{fig:insert_example}~(a)) for $interval_{n-1}$, we continue to search for the object insertion coordinates under the remaining ratio intervals. Instead of utilizing the strategy in \textit{step-1} to find a new coordinate, which is time-consuming, we search the coordinate along a line. We denote the centroid coordinate of the largest object in background image as $(x_{max}, y_{max})$, which is point $A$ in Fig.~\ref{fig:insert_example}~(a) (line 17). We construct a line $L$ (red lines in Fig.~\ref{fig:insert_example}) by connecting point $(x_{max}, y_{max})$ and point $(x_{n-1}, y_{n-1})$ and extend the line to point $(x_{bound}, y_{bound})$ (point $B$ in Fig.~\ref{fig:insert_example} (a)), where $(x_{bound}, y_{bound})$ is the farthest coordinate on $L$ keeping the inserted object inside the background image. 
{\methodname} then employs binary search to find a suitable coordinate that obeys the rules $R_1$ and $R_2$ along the line (line 18).
Fig.~\ref{fig:insert_example} (b), (c), (d) demonstrate an example of the binary search process along the line when {\methodname} intends to find a suitable coordinate for ratio interval (0, 0.15]. In the first search, it gets an overlapping ratio that equals to 0\% which is too small. Then it performs the second search and gets the overlapping ratio that equals to 17\%, which violates $R1$. 
Finally, we find coordinate $(x_1, y_1)$ that satisfies $R_1$ and $R_2$ for the ratio interval. We could see that the saliency of the cat is not affected by the inserted dog after insertion. 
For each ratio interval, we allows at most $c_2$ "jumps" in the binary search, where $c_2$ is a pre-defined threshold.
If we cannot find a suitable position for the interval, we will randomly select a new image and starts from \textit{step-1} again.



\begin{algorithm}[h]
	\small
	\caption{ An implementation of location tuning }
	\label{alg:location}
	\begin{flushleft}
		\textbf{Input:} the background $b$, a list of object candidates $objects$, number of overlapping ratios $n$, the max overlapping ratio $ratio_{max}$, resizing interval parameters $\alpha$ and $\beta$, patience parameters $c_1$ and $c_2$  \\
		\textbf{Output:} a size for the inserted object, and a list of locations for controlling different overlapping ratios
	\end{flushleft}
	\begin{algorithmic}[1]
		\State $location\_list\leftarrow List(\hspace{0.5ex})$
		\Comment{Initialize with empty list}	
		\For {$obj$ in $objects$}
		\State $coordinate_{max} \gets$  \Call{RandomCoordinate}{ } 
		\State $ObjNewSize \gets$ \Call{RandomSize}{$\alpha$, $\beta$, $obj$, $b$}
		\State $obj \gets$ \Call{Resize}{$obj$, $ObjNewSize$} 
		
		\State $patience_1 \leftarrow$ $0$
		\While {$R_1$ is False or $R_2$ is False for $ratio_{max}$}
		\Comment{Step 1}
		\If{$patience > c_1$ }
		\State break
		\EndIf
		\State $coordinate_{max} \gets$  \Call{RandomCoordinate}{ }
		\State $patience_1+=1$
		\EndWhile

		\For {i in range(0,n-2)} 
		\Comment{Step 2}
		\State $patience_2 \leftarrow$ $0$
		\While {$R_1$ is False or $R_2$ is False for $ratio_{i}$}
		\If{$patience > c_2$ }
		\State break
		\EndIf
		\State $centroid \gets $ \Call{Coordinate}{LargestObjOf(b)}
		\State $coordinate_{i} \gets$ \Call{BinarySearch}{$centroid$, Boundary(b)}
		\State $patience_2+=1$
		\EndWhile
		\State $location\_list$.append($coordinate_{i}$)
		\EndFor
		\If {$Length(location\_list)==n$}
		\State break
		\EndIf
		\EndFor
		\State \Return $ObjNewSize$, $location\_list$
		
	\end{algorithmic}
\end{algorithm}

\subsection{Caption Collection}
After object insertion, {\methodname} constructs image pairs, where each pair contains the background image and a corresponding synthetic image. These images will be input to the IC system under test and the returned captions will be collected. In this paper, we test one paid IC service, \textit{i.e.}, Microsoft Azure Cognitive Services~\cite{miocrosoft_azure_api}, and five popular IC models, \textit{i.e.}, Attention~\cite{xu_show_2015}, $Oscar_B$, $Oscar_L$~\cite{li_oscar_2020}, $VinVL_B$, $VinVL_L$~\cite{zhang_vinvl_2021}. For the paid IC service, we invoke the API provided by Microsoft Azure Cognitive Services~\cite{miocrosoft_azure_api}. For the IC models, we use the open-source code provided by the authors and train our own IC models following the exact description in the original papers.

\subsection{Error Detection}
After caption collection, {\methodname} inspects the captions of every image pair and reports them as a suspicious issue if they violate our metamorphic relations (MRs). 
{\methodname} provides two MRs. 

\textbf{MR1.} We denote the IC system as $I$, the background image as $b$, the generated image as $i'$, the inserted object as $obj$, the caption pair produced by $I$ are $I(b)$ and $I(i')$.  \textit{MR1} is defined as follows:
\begin{equation}
    Set(I(i')) == Set(I(b)) \cup \{obj\},
\end{equation}
where $Set(I(i'))$ denotes the set of object classes in $I(i')$, $Set(I(b))$ indicates the set of object classes in $I(b)$. 

\textbf{MR2.} We use a function $F_*(y)$ to denote the singular-plural form of an object class $y$ in caption $I(*)$, indicating whether the object class $y$'s form is singular or plural. For example, $F_b(dog)$ is the singular-plural form of ``dog" class in the caption of background image $b$.
$C$ denotes the object classes in both $I(b)$ and $I(i')$; $c_k$ is the $k$-th object class in $C$. If the inserted object $obj$ is not in $Set(I(b))$
, \textit{MR2} requires:
\begin{equation}
    \begin{aligned}
    F_b(c_k) = F_{i'}(c_k), \\
    F_{i'}(obj) = singular.
    \end{aligned}
\end{equation}
If the inserted object $obj$ is in $Set(I(b))$, we consider $C'=C\setminus\{obj\}$, where $c_j$ is the $j$-th object class in $C'$. Then the second MR requires:
\begin{equation}
    \begin{aligned}
    F_b(c_j) = F_{i'}(c_j), \\
    F_{i'}(obj) = plural.
    \end{aligned}
\end{equation}
If any of the two MRs are violated, the original image, the generated image, and the corresponding captions will be reported as a suspicious issue.



In our implementation, we mark up the word in caption according to a particular part of speech (POS) via a POS tagging tool, \textit{i.e.}, XPOS~\cite{marcus1993building}. 
We extract words from the captions whose POS are NN or NNS, corresponding to singular noun or plural noun, respectively. Then we can obtain the a set containing all the object classes in a caption, and construct the mapping for function $F$ which assigns the singular-plural form to the object class in the set. 

\section{Evaluation}\label{sec:exper}

\subsection{Experimental Setup and Dataset}

\begin{table*}[ht]
		\centering
		\caption{Precision (true positives/suspicious issues) of Deeptest~\cite{tian_deeptest_2018} and {\methodname} on one paid API and five models.}
		\includegraphics[width=0.95\linewidth]{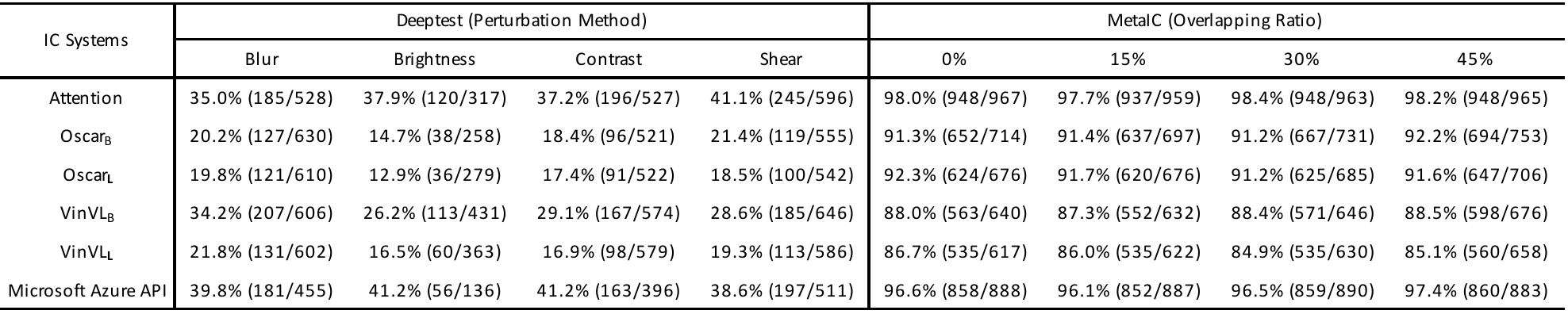}
		\label{fig:table_precision}
\end{table*}

\subsubsection{Experimental environments}
 All experiments are run on a Linux workstation with an 8 core AMD Ryzen 5800X 4.8GHz Processor, 64GB DDR4 2666MHz Memory, and GeForce RTX 3090 GPU. The Linux workstation is running 64-bit Ubuntu 20.04.2 LTS with Linux Kernel 5.11.0. For POS tagging, we use the XPOS implemented in Stanza,\footnote{https://stanfordnlp.github.io/stanza/} an NLP Package powered by Stanford NLP Group.

\subsubsection{Dataset}
To show that {\methodname} can work effectively on different image sources, we collect object source images from Flickr~\cite{flickr} and background images from MS COCO Caption~\cite{chen2015microsoft}, a standard dataset in image captioning field. For MS COCO Caption Dataset, we focus on the images whose class names are single words (60 out of 80) because the POS tagging technique can only assign POS to single words. 
It only requires decent engineering effort to generalize {\methodname} to class names containing multiple words. All the code and datasets in this paper will be open-source.

\subsection{Precision}
{\methodname} automatically reports suspicious issues, where each issue contains a pair of images and their captions $(I(b), I(i'))$. 
Therefore, the effectiveness lies in how precise the reported issues are. In this section, we try to answer the following question: how many of the reported issues contain real captioning errors. We test one well-known IC API: Microsoft Azure API, and five IC models: Attention~\cite{xu_show_2015}, $Oscar_B$, $Oscar_L$~\cite{li_oscar_2020}, $VinVL_B$, $VinVL_L$~\cite{zhang_vinvl_2021}. In this experiment, we set $ratio_{max}=45\%$ and $n=4$ to balance the diversity of the test cases (in terms of the overlapping ratio) and the test case quality. We believe this parameter setting can be used in other datasets in general because the MS COCO Caption contains diverse images (different sizes and salient objects). 
We use {\methodname} to randomly generate 1,000 background-object pairs from MS COCO Caption and our object corpus, and construct 1,000 synthesized images for every overlapping ratio interval.
After synthesizing the images, we obtain 4,000 image pairs $(b, i')$. We collect 4,000 caption pairs $(I(b), I(i'))$ from each of the IC systems under test. Based on these caption pairs, {\methodname} returns a list of suspicious issues. 

We are the first to test IC systems so there are no available baselines. Intuitively, the metamorphic testing techniques for simpler tasks that also take images as input could be adapted to testing IC systems. Thus, we adopt Deeptest~\cite{tian_deeptest_2018}, a metamorphic testing technique for image classifier in our precision experiment. Specifically, we use four image transformations in Deeptest (blur, brightness, contrast, shear) and check whether the image pair share the same caption. We use the same background images as the experiment setting of {\methodname} to synthesize 4,000 image pairs $(b, i')$, and collect 4,000 caption pairs $(I(b), I(i'))$ from the IC systems.

To verify the results, two authors manually inspect all the suspicious issues separately following the instructions on data labeling in MS COCO paper~\cite{chen2015microsoft}, including "Describe all the important parts of the scene", "Do not describe things that might have happened
in the future or past", \textit{etc.} During this manual analysis, all disagreements were discussed until a consensus was reached. The results of this phase have a Cohen's kappa of 0.822, showing a substantial-level of agreement~\cite{mchugh2012interrater}.


\subsubsection{Evaluation Metric}
If the caption pair $p=(I(b), I(i'))$ is reported as suspicious issue, and $I(b)$ or/and $I(i')$ contains captioning error(s), then we set $error(p)$ to be true, otherwise we set $error(p)$ to be false. Given a list of suspicious issues, the precision is calculated by:
\begin{equation}
    Precision = \frac{\sum_{p\in P}{error(p)}}{|P|},
\end{equation}
where $P$ are the suspicious issues reported by {\methodname} and $|P|$ is the number of the suspicious issues.




\subsubsection{Results}
The results are presented in Table~\ref{fig:table_precision}, the precision of {\methodname} and Deeptest~\cite{tian_deeptest_2018} on one API and five models. 
The precision of {\methodname} ranges from 84.9\% to 98.4\%, while Deeptest's precision ranges from 12.9\% to 41.2\%. 
Specifically, {\methodname} achieves a precision of (96.1\%-97.4\%) on Microsoft Azure API, and a precision of (84.9\%-98.4\%) on the five IC models.
The results demonstrate that {\methodname} is much more precise in testing IC systems than existing metamorphic testing techniques for systems that also take images as input.
In addition, {\methodname} reports much more erroneous issues. Specifically, {\methodname} reports 535 to 948 erroneous issues, while Deeptest reports 36 to 245.
{\methodname} successfully report 16,825 erroneous issues, which contains a total number of 17,380 captioning errors.
Specifically, out of the 16,825 erroneous issues reported, 120 issues are reported because of captioning errors in the original images.
Note that the comparison between {\methodname} and Deeptest is not apple-to-apple as Deeptest was originally designed for testing image classifiers.
We regards Deeptest as a baseline here for the completeness of discussion.
The images synthesized by {\methodname} are diverse.
From the table, we can observe that {\methodname} consistently achieves high precision under different overlapping ratios, showing the effectiveness of our object resizing and location tuning algorithms. 
For example, when we use {\methodname} to test the Microsoft Azure API~\cite{miocrosoft_azure_api}, the precision values for $\{ratio_0, ratio_1, ratio_2, ratio_3\}$ are $\{96.6\%, 96.1\%, 96.5\%, 97.4\%\}$, respectively. 

\subsubsection{False positives}
\begin{figure}[ht]
		\centering
		\includegraphics[width=1.0\linewidth]{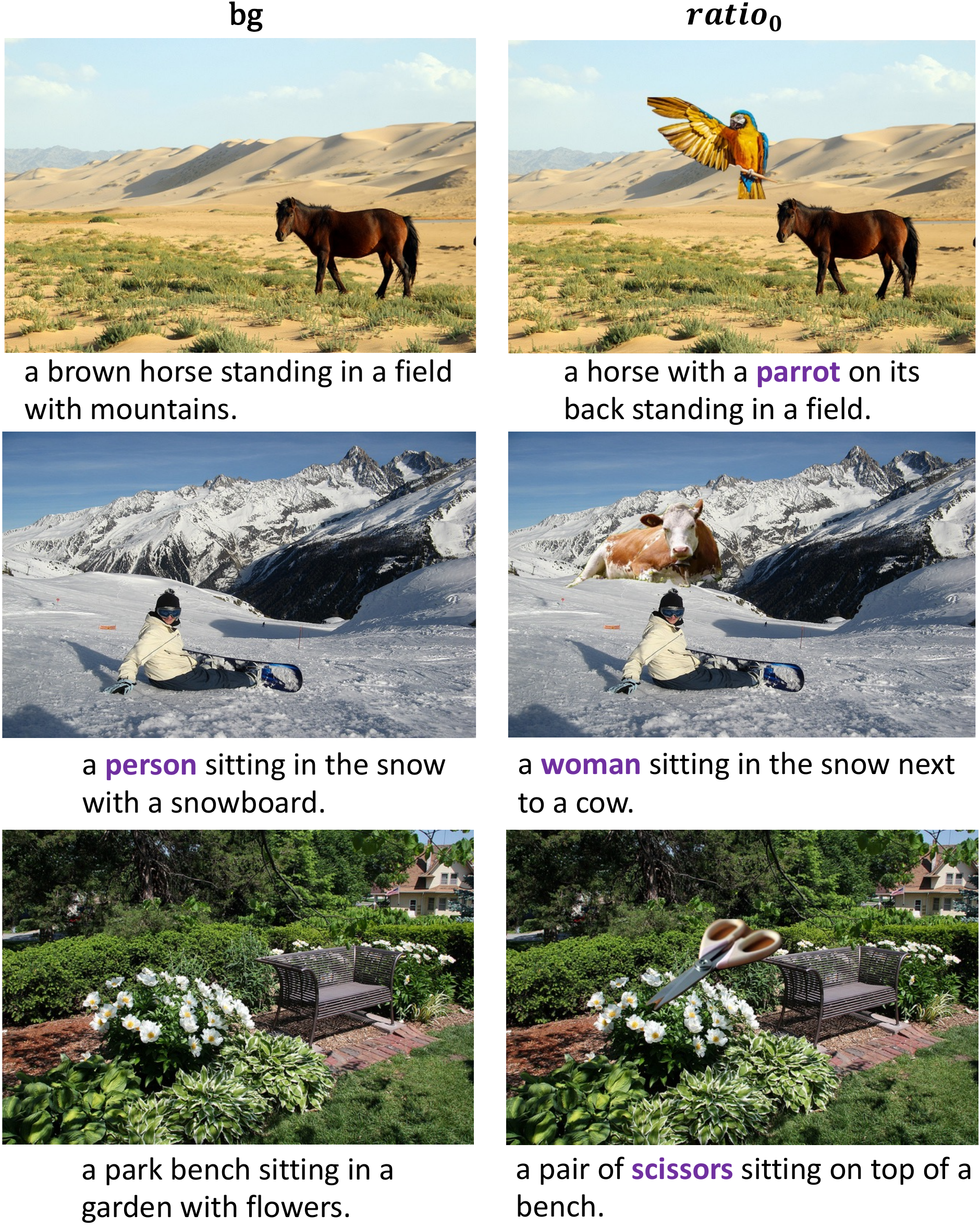}
		\caption{False positives reported by {\methodname}.}
		\label{fig:false_positive}
\end{figure}

To further understand the results, we categorize the false positives of {\methodname} and discuss the prospective solutions for them.
Although {\methodname} achieves very high precision in our evaluation, we still encounter some false positives. We manually inspect the false positives and present three typical examples in Fig.~\ref{fig:false_positive}. 
First, after inserting a bird into the background image, the caption describes the "bird" as "parrot", which is correct because parrots are subspecies of birds. 
However, our class set does not contain "parrot", leading to a false positive. Second, after inserting a cow into the background image, the depiction of the person changes from "person" to "woman". A dictionary containing more object types can effectively reduce these two kinds of false positives.
Third, "a pair of scissors" is regarded as an incorrect singular-plural form of "scissors" by our method, while in modern English, the word "scissors" has no singular form, leading to the false positive. In the future, adopting a more advanced POS tagging tool or maintaining a dictionary of words of special singular-plural form can help.

\begin{figure*}[ht]
		\centering
	\includegraphics[width=0.94\linewidth]{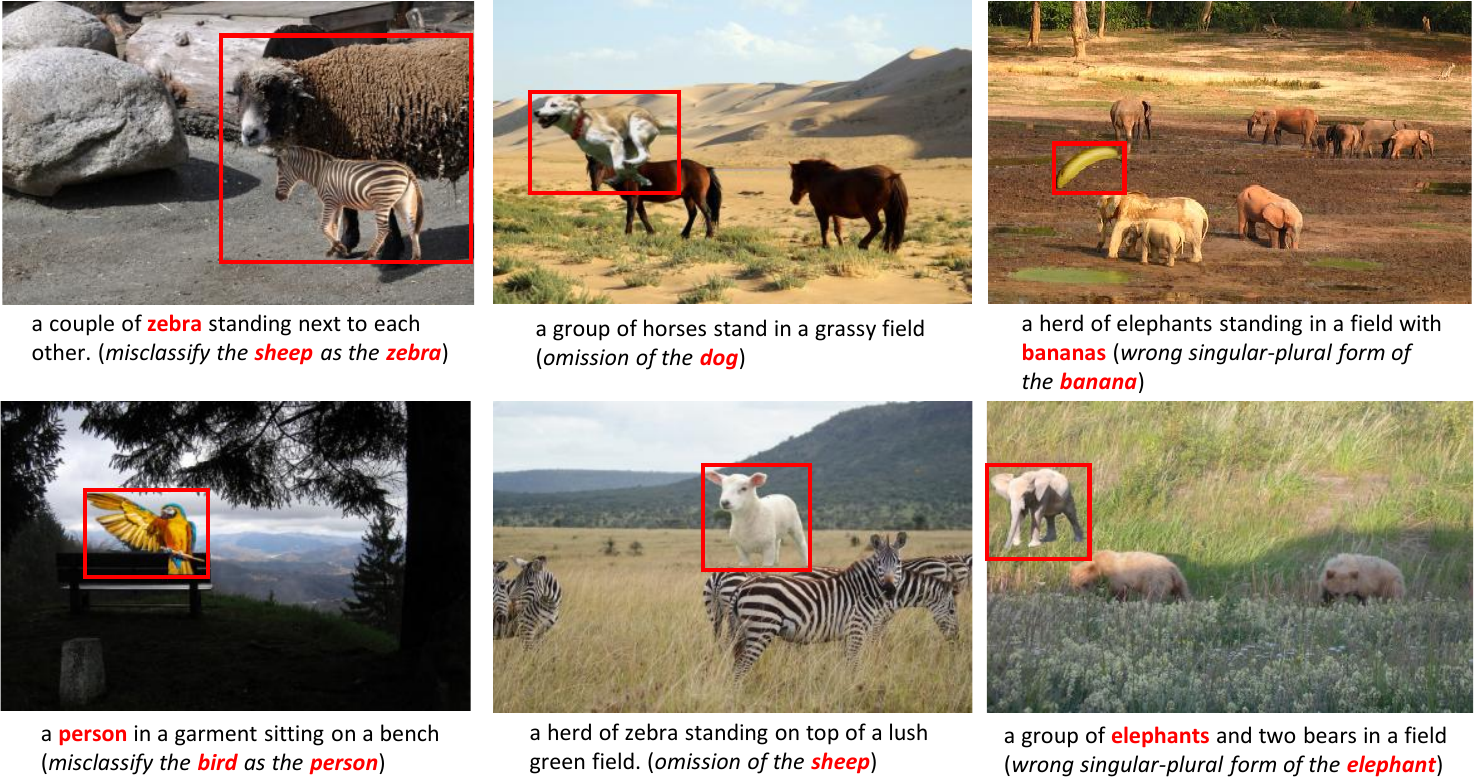}
	\caption{Examples of captioning errors reported by {\methodname}.}
		\label{fig:error_cases}
	\vspace{4ex}
\end{figure*}

\begin{table*}[ht]
		\centering
		\caption{Ablation study of the metamorphic relations.}
		\includegraphics[width=0.99\linewidth]{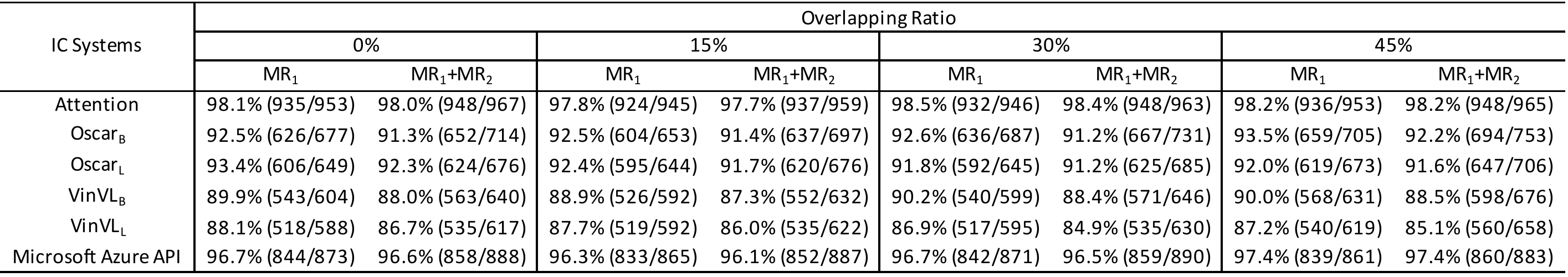}
		\label{fig:table_ablation}
\end{table*}

\subsection{Erroneous Captions}
{\methodname} is capable of finding three kinds of captioning errors: 
\begin{itemize}
  \item \textbf{Classification error} is an error that IC systems provide an incorrect class name for an object, for instance, an erroneous caption depicts an airplane as a skateboard.
  \item \textbf{Recognition error} is an error that IC systems omit the description of a salient object in the image.
  \item \textbf{Single-plural error} is an error that IC systems return the incorrect singular-plural form of an noun in the caption. For example, an erroneous caption gives a depiction "bananas" when there is only one banana in the image.
\end{itemize}

To provide a glimpse of the diversity of the uncovered errors, this section highlights examples for all the three kinds of errors in Fig.~\ref{fig:error_cases}.
The first column corresponds to the classification errors, \textit{e.g.}, the caption in the first row uses "a couple of zebra" to describe the image including a sheep and a zebra, and the caption of the second row incorrectly describes the bird as a person in a garment. The second column corresponds to the recognition errors, where the caption of the first row misses the depiction of a dog near two horses, and the caption of the second row omits the sheep in a group of zebras. The third column shows the errors of singular-plural form, where the caption of the first row describes a single banana as bananas, and the caption of the second row depicts a single elephant as two elephants.

\subsection{Ablation Study}
We conduct the ablation study to evaluate the effectiveness of the two metamorphic relations we propose.
From Table~\ref{fig:table_ablation} we can see that the precision of $MR_1$ and $MR_1+MR_2$ is very close to each other in most cases.
Sometimes, the precision will be slightly decreased when we use $MR_1+MR_2$, while we will report more erroneous issues.
For example, for the $Oscar_B$ model with 45\% overlapping ratio, precision is decreased from 93.5\% to 92.2\%.
However, with $MR_1+MR_2$, we can find 35 more erroneous issues than only using $MR_1$, which shows that $MR_2$ is useful and necessary.

\subsection{Case Study on IC Errors via Visualization}
To better understand the reported captioning errors and explore the potential root causes, in this section, we conduct visualization experiments. Specifically, we try to answer the question: "why the IC systems return incorrect captions given the synthesized images?" We visualize the errors in both CNN-RNN-Based IC systems (via attention of the generated object words) and VLP IC systems (via the prediction of faster R-CNN). 

\subsubsection{CNN-RNN-Based IC}
For CNN-RNN-Based IC, we explore errors in Attention~\cite{xu_show_2015} because it achieves high accuracy among all the CNN-RNN-Based ICs. We visualize the attention mechanism as it is the most important component of the model. 
Specifically, we intend to show how different overlapping ratios would affect the attention mechanism and the caption.

\begin{figure*}[ht]
		\centering		\includegraphics[width=0.95\linewidth]{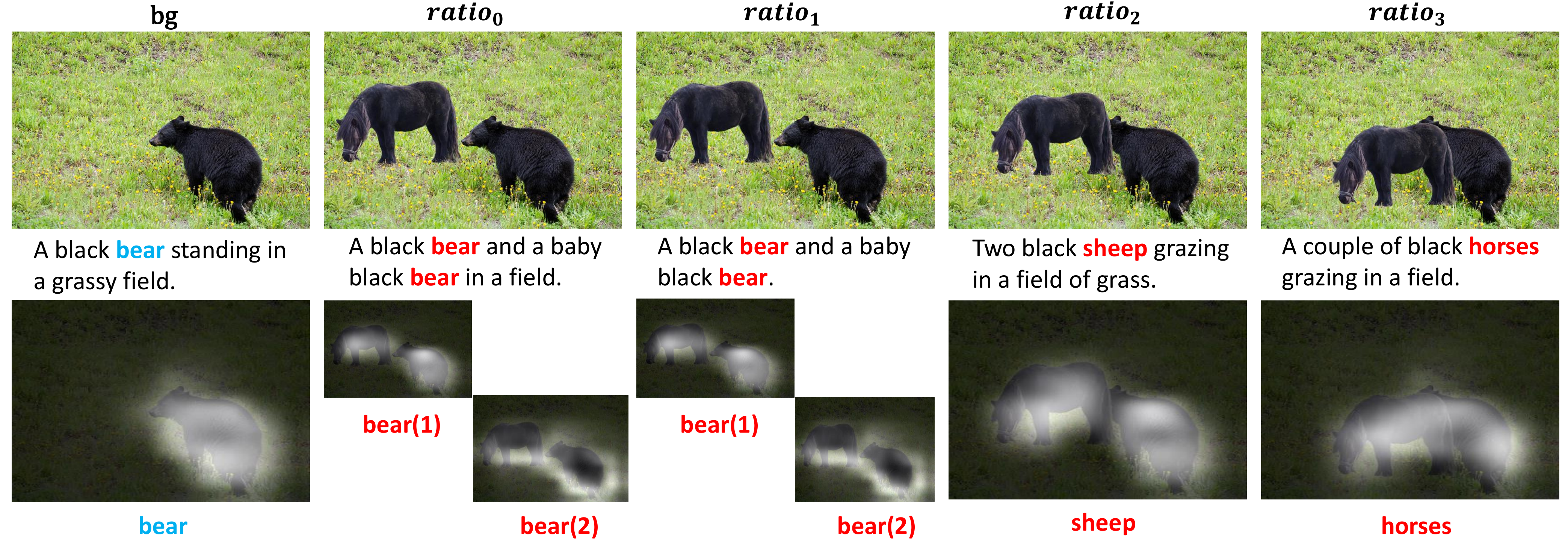}
		\caption{Attention mechanism fails to depict the images.}
		\label{fig:show_bear_horse}
\end{figure*}

In Fig.~\ref{fig:show_bear_horse}, we present the attention mask changes after inserting an object to the original background image. 
We could observe that, at the beginning, the black bear is correctly captioned in the background image. However, when we insert the horse into the background image with different overlapping ratios ($n=4$, $ratio_{max}=45\%$), we find that the captions become incorrect. For $ratio_0$ and $ratio_1$, the black horse is misclassified as a black bear. For $ratio_2$, the black horse and black bear have been both recognized as black sheep. For $ratio_3$, the black bear is misclassified as black horse. The second row of Fig.~\ref{fig:show_bear_horse} shows the attention masks generated by the model, which serves as an important component in its captioning process. By observing the attention masks of $ratio_0$ and $ratio_1$, we can see that the attention masks for the two words of "bear" in the caption mistakenly highlight the regions of both the horse and the bear. This indicates that the captioning error is likely to be caused by model's attending to incorrect region. Thus, research on improving the attention mechanism in demand.

\begin{figure*}[ht]
		\centering
		\includegraphics[width=1.0\linewidth]{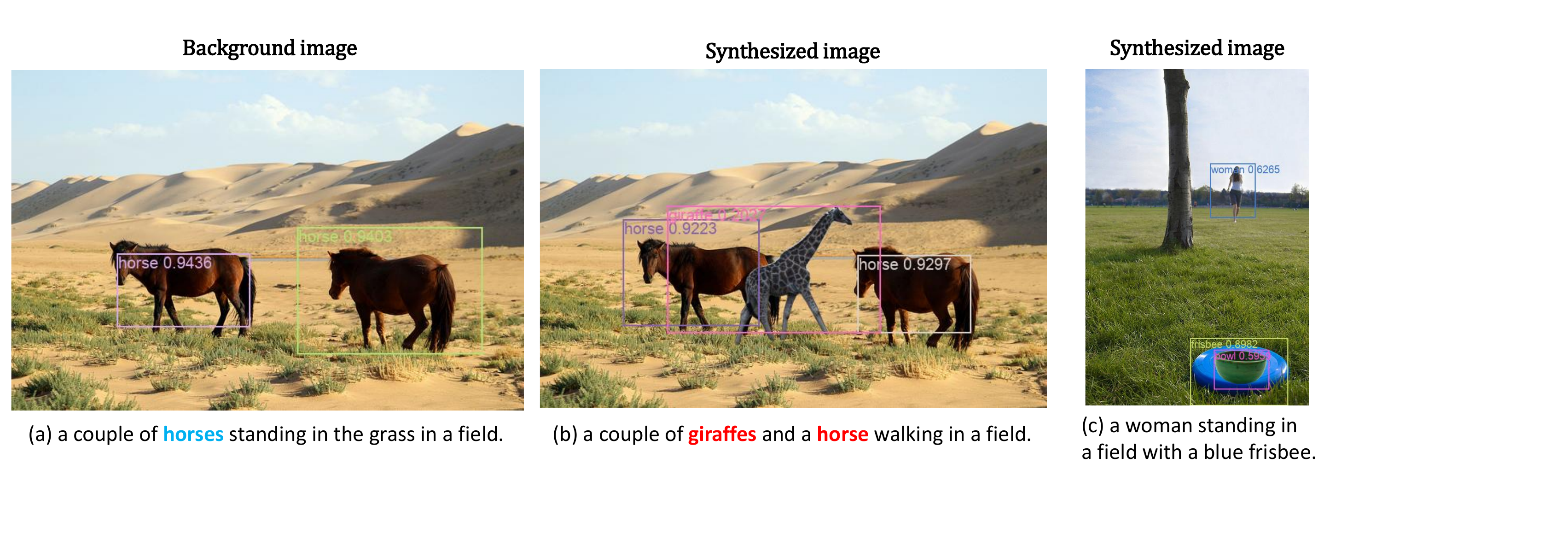}
		\caption{Errors caused by the first component (a)(b) and the second component (c) in VLP IC.}
		\label{fig:vlp_fe_errors}
\end{figure*}

\subsubsection{Perturbation to VLP IC}
{\methodname} also reports many erroneous captions produced by VLP ICs. Specifically, we choose a typical VLP IC model $Oscar_B$ and conduct visualization experiment to show the vulnerabilities of the two components of VLP ICs, including Faster R-CNN in the first stage and BERT in the second stage. 
By comparing Fig.~\ref{fig:vlp_fe_errors} (a) with Fig.~\ref{fig:vlp_fe_errors} (b), we show an example where its erroneous caption is caused by Faster R-CNN (the first component). 
We draw the bounding boxes produced by the Faster R-CNN of $Oscar_B$ for the background and the synthesized image. 
For clarity, we only draw the bounding boxes for the object in the image. 
The caption produced by $Oscar_B$ is correct for the background image.
However, for the synthesized image, the caption describes a single giraffe as "a couple of giraffes", resulting in an error of singular-plural form. 
We could observe that the bounding boxes for background image are of high accuracy. 
However, the bounding box of the giraffe in the synthesized image and that of the horses have a big overlap. 
The part of the image framed by the bounding box will be encoded into region-feature, which acts as the major input to the second component of $Oscar_B$.
The second component uses region features and the corresponding tags (\textit{e.g.}, "giraffe") to generate the caption.
Specifically, $Oscar_B$ concatenates the tag "giraffe" with the region-feature for cross-modal representation learning.
We think the erroneous depiction of "a couple of giraffes" could be caused by the low-quality region-feature of "giraffe" (\textit{i.e.}, the sub-optimal bounding box). 
It overlaps a lot with the horses, and thus $Oscar_B$ captions one of the horses as a giraffe. 


The visualization experiment also shows that the erroneous captions may be caused by the glitches in the BERT module the second component of VLP ICs. 
As shown in Fig.~\ref{fig:vlp_fe_errors} (c), the bounding boxes of the objects in the image are of high accuracy, while the generated sentence omits the description of the bowl on the frisbee. 
The visualization case study indicates that {\methodname} can report captioning errors cause by both the first component (image) and the second component (text).

\subsection{Finding Labeling Errors in the Training Corpus}

Labeling errors in the training corpus has been a prevalent and severe problem in machine learning and deep learning. Even the most famous Datasets have errors in their labels, such as ImageNet~\cite{krizhevsky2012imagenet}, CIFAR~\cite{krizhevsky2009learning}, and MNIST~\cite{deng2012mnist}. Several methods~\cite{sambasivan2021everyone,zhang2017improving} have been proposed to tackle this problem, mainly focusing on noises in the data. 
Recently, AI software applications are approaching human-level performance.
For IC task, Hu \textit{et al.}~\cite{hu_vivo_2020} claims that their model has achieved a better result than human in terms of CIDEr~\cite{vedantam2015cider} score in Nocaps~\cite{agrawal2019nocaps} task. 
Given that these models have been well trained on the existing corpora, fixing labeling errors in these corpora is a reasonable way to further improve their performance. 
In this section, we explore whether it is possible to adapt the high-level idea of {\methodname} detecting labelling errors in a standard dataset MS COCO Caption~\cite{chen2015microsoft}.

Specifically, denote $GT(b)$ as the label caption of image $b$ in MS COCO Caption, we have a tuple $(I(b), I(i'), GT(b))$.
Denote the common object class set between $I(b)$ and $I(i')$ as $Set_{invar}$:
\begin{equation}
    Set_{invar} = Set(I(b)) \cap Set(I(i')),
\end{equation}
Intuitively, objects classes in both $(I(b)$ and $I(i')$ imply the important objects in the image. We denote the object class set of $GT(b)$ as $Set(GT(b))$, it should satisfy the relation that:
\begin{equation}
    Set_{invar} \subset Set(GT(b)).
\end{equation}
Otherwise, $GT(b)$ may depict image $b$ improperly since it is likely to miss important objects in its caption. 
We map the object class to their super-category defined in MS COCO Caption and use the super-category in the aforementioned method.

\begin{figure}[ht]
		\centering
		\includegraphics[width=1\linewidth]{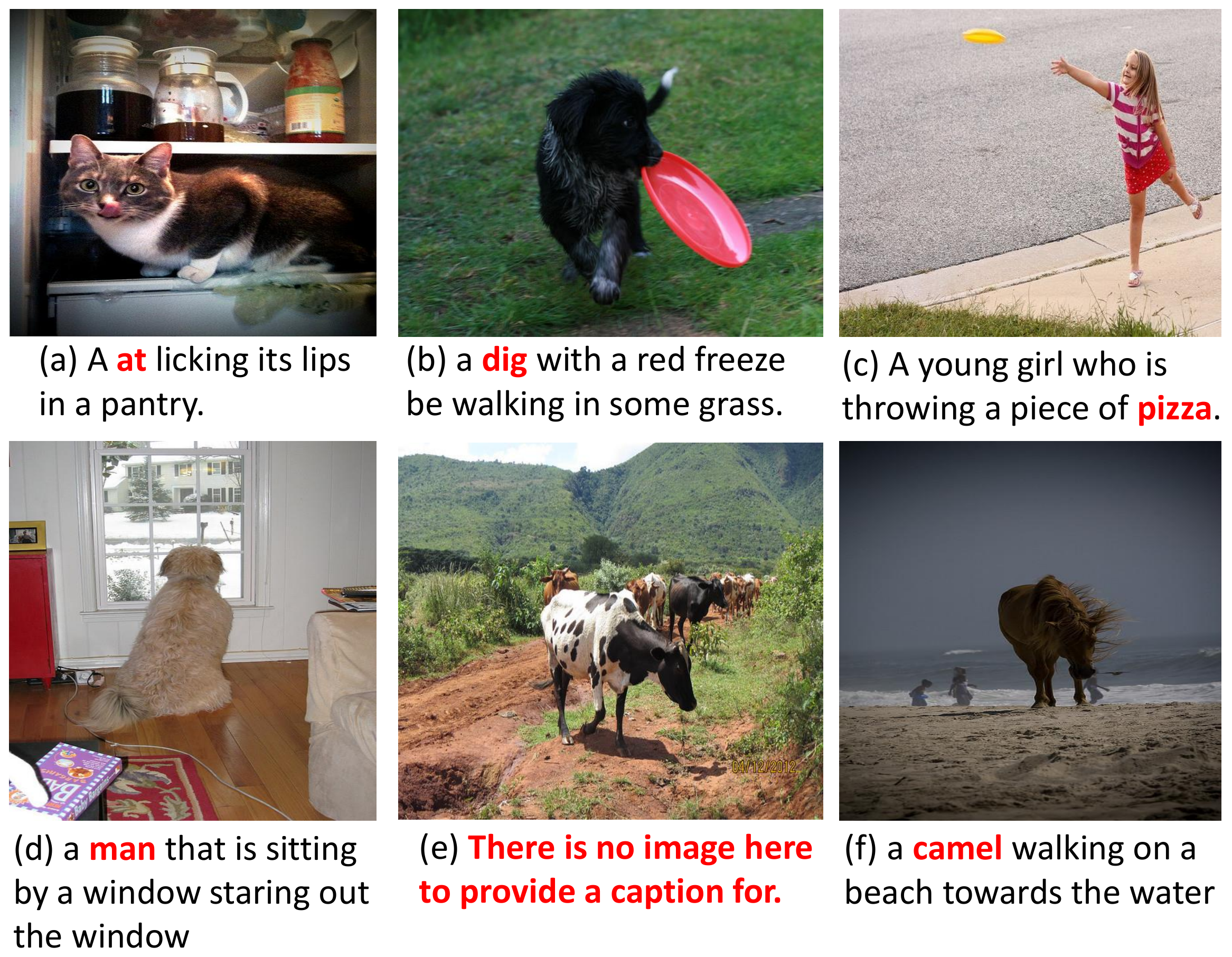}
		\caption{Labeling errors in MS COCO Caption~\cite{chen2015microsoft} reported by {\methodname}.}
		\label{fig:gt_error_cases}
\end{figure}

We choose ten object classes $\{dog, cat, sheep, truck, cow, zebra,\\ elephant, horse, frisbee, bird\}$, construct 6,662 tuples of $(I(b), I(i'), \\GT(b))$, and try to find erroneous labels of these classes in MS COCO Caption. 
By using this technique adapted from {\methodname}, we have found 151 caption errors, using only a small portion of the data in MS COCO Caption.
We present examples of incorrect label errors in Fig.~\ref{fig:gt_error_cases}, including typos, misclassification error, "no-image" error, \textit{etc.} 
For example, caption (a) in the first row mistakenly writes "at" rather than "cat"; caption (c) mistakenly describe a "frisbee" as a "pizza"; caption (e) is a complain message rather than a caption; and caption (d) incorrectly depicts a "dog" as a "man". 

Given that the performance of deep learning models mainly depends on the model structure and the parameters learned from the training dataset, we believe the labeling errors we find with {\methodname} indicate the usefulness and wide applicability in enhancing the robustness of IC software.
Thus, it reflects the practical relevance of {\methodname}. The 151 labeling errors are found from 6,662 captioned images in the training data.
We believe our approach can detect more errors on a larger input set. 
Although finding labeling errors is not the main focus of this paper, we think the experiment demonstrates the wide-applicability of {\methodname}.

\subsection{Retraining with Erroneous Issues}
To explore whether the erroneous issues reported by {\methodname} can be utilized to improve IC systems, we re-label the erroneous issues following the labeling standard of MS COCO Caption~\cite{chen2015microsoft}, and fine-tune the $Oscar_B$ model. 
Specifically, we add the re-labeled synthesized images to MS COCO Caption and perform fine-tuning for 40K steps. 
We follow the publicly available splits\footnote{https://cs.stanford.edu/people/karpathy/deepimagesent/} as in the original Oscar paper~\cite{li_oscar_2020} and use our synthesized images as the augmented data in the training set.

Table~\ref{fig:table_finetune} presents the results. 
We can observe that BLEU-4 increases by 1.2\%, METEOR increases by 3.7\%, CIDEr increases by 2.5\%, and SPICE increases by 8.8\%. Specifically, the SPICE~\cite{anderson2016spice} has a system-level correlation of 0.88 with human judgements on MS COCO~\cite{lin2014microsoft}. 
According to a survey of IC systems (Table 2 in \cite{stefanini2021show}), our improvement of the four score are significant.
In addition, we can see that except for BLEU-4, the other scores of $Oscar_{B-Fintune}$ are higher than $Oscar_L$.
The fine-tuned model achieves competitive performance with much fewer parameters and less training time.
The results indicate that the captioning errors in the synthetic images can be utilized to further improve the performance of IC systems, both in efficiency and accuracy, demonstrating its practical relevance.
Note that although improving IC system using test cases is an interesting and important topic, it is not the focus of this paper. Thus, we regard it as a promising future direction.

\begin{table}[ht]
		\centering
		\caption{Performance of fine-tuned model.}
		\includegraphics[width=1\linewidth]{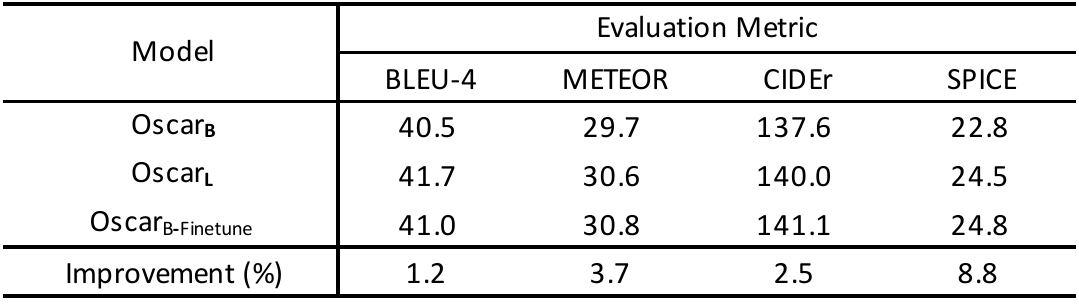}
		\label{fig:table_finetune}
\end{table}



\section{Related Work}\label{sec:related}


\subsection{Robust AI Software}
In recent years, deep neural networks have achieved great success in a variety of fields and thus artificial intelligence (AI) software has been widely used in our daily lives. However, AI software can generate erroneous outputs that cause severe accidents and even endanger the users~\cite{accident1,accident2,accident3}. To explore the vulnerability of AI software, a line of research has focused on attacking various AI systems, such as classification systems~\cite{goodfellow2014explaining, luo2018towards, zhao2020towards}, and automatic speech recognition systems~\cite{qin2019imperceptible, carlini2018audio}. These papers fool the AI software with imperceptible perturbations. Meanwhile, to enhance the reliability of AI software, a variety of related topics have been studied, including testing~\cite{du2019deepstellar, gambi2019automatically, henriksson2019towards, he2020structure,he2021testing,Kang2020aa}, online adversarial detection~\cite{ma2019nic, tao2018attacks, wang2019adversarial, gu2019detecting}, and robust training mechanisms for deep neural networks~\cite{kannan2018adversarial, lin2019defensive, madry2017towards, papernot2016distillation}. Most of these methods are white-box, largely dependent on the knowledge of the model internals; while our approach is black-box, which can be easily adapted to test any IC systems.


\subsection{Multimodal Task and Image   Captioning}

Modality refers to the way in which something happens or being sensed by human. Accordingly, a multimodal task indicates that the task involves more than one modality (\textit{e.g.}, image and text for image captioning). The corresponding AI models need to learn these multimodal signals collectively, which brings new challenges as introduced by Baltru{\v{s}}aitis et al.~\cite{baltruvsaitis2018multimodal}: representation, translation, alignment, fusion, and co-learning. Typical multimodal tasks include speech recognition and synthesis~\cite{hannun2014deep, kalchbrenner2018efficient}, cross-modal retreival~\cite{gao2020fashionbert, vo2019composing}, and image captioning~\cite{xu_show_2015, li_oscar_2020, zhang_vinvl_2021, vinyals_show_2015}. Compared with single-modal systems (\textit{e.g.}, image classifier or machine translation), multimodal systems are more difficult to test because we need to consider the translation from one modality to another and most of the existing testing techniques for single-modal models cannot be used here without non-trivial adaptations.

Recent years have witnessed the blossom of IC systems, where researchers focus on improving the accuracy of IC. For CNN-RNN-Based IC, researchers designed various CNNs and RNNs to enhance the intermediate representation~\cite{karpathy2015deep,vinyals_show_2015,xu_show_2015}. For VLP IC, researchers proposed multiple pre-trained techniques~\cite{zhou2020unified, li_oscar_2020, hu_vivo_2020, zhang_vinvl_2021}. Different from these papers that make effort to achieve high accuracy, {\methodname} aims to improve the robustness of IC. To this end, several papers focus on attacking existing IC models in a white-box manner~\cite{xu2019exact, chen2018attacking, xu2018fooling}, which requires the complete knowledge of underlying networks. Differently, {\methodname} is black-box, which does not rely on model internals, such as network structure and parameters.


\subsection{Metamorphic Testing}
Metamorphic Testing (MT) is a general methodology that applies a transformation to test input(s) and observes how the program output turns into a different one as a result~\cite{chen2020metamorphic, segura2016survey, chen2018metamorphic}. The core idea of MT is to verify the MRs between the outputs from multiple runs of the program with different inputs, which is useful when test oracle is lacking. MT is widely adopted in traditional software, such as compilers~\cite{le2014compiler,lidbury2015many}, datalog engines~\cite{mansur2021metamorphic}, and service-oriented applications~\cite{chan2005towards, chan2007metamorphic}. Recently, it has also been used in testing AI software, such as autonomous cars~\cite{tian_deeptest_2018, zhang2018deeproad}, statistical classifiers~\cite{xie2009application, xie2011testing}, object detection~\cite{wang2020metamorphic}, and machine translation~\cite{he2020structure, he2021testing}. In this paper, we propose {\methodname}, a novel, widely-applicable metamorphic testing approach for image captioning.

\section{Conclusion}\label{sec:con}

In this paper, we propose the first black-box testing approach, {\methodname}, for validating image captioning systems. The distinct benefits of {\methodname} are its simplicity and generality, and thus wide applicability. {\methodname} can effectively disclose many captioning errors. In our experiments, {\methodname} successfully reports 16,825 erroneous issues in six IC systems with high precision (84.9\%-98.4\%) revealing 17,380 captioning errors. In addition to the main focus of this paper (\textit{i.e.}, testing), to further understand the reported errors, we visualize the attention of objects in CNN-RNN-Based ICs and the predictions of Faster R-CNN used in VLP ICs, which explores the major root causes behind. We also show that {\methodname} can be adapted to find errors in the standard dataset of image captioning, indicating its potential in further enhancing the reliability of IC systems. For future work, we plan to extend {\methodname} by considering other kinds of caption constituents (\textit{e.g.}, verbs). We will also explore automated error detection for standard datasets, which we regard as an important future direction.

\begin{acks}
We thank the anonymous ISSTA reviewers for their valuable feedback on the earlier draft of this paper. This paper was supported by the National Natural Science Foundation of China (No. 62102340). Part of the experiments were conducted on the high-performance computing platform, which is managed by the Information Technology Services Office (ITSO) at the Chinese University of Hong Kong, Shenzhen.
\end{acks}



\balance
\bibliographystyle{ACM-Reference-Format}

\bibliography{References}



\end{document}